\documentclass[preprint, superscriptaddress, nofootinbib,preprintnumber]{revtex4-2}

\usepackage[colorlinks=true, linkcolor=blue, citecolor=blue, urlcolor=blue]{hyperref}
\usepackage{amsmath, graphicx}
\usepackage{subfigure}
\usepackage{xcolor}
\usepackage{braket}
\usepackage{array}
\usepackage[most]{tcolorbox}
\usepackage[normalem]{ulem}

\begin{document}

\title{
Scattering Entanglement Entropy and Its Implications for Electroweak Phase Transitions
}

\author{Jia Liu}
\email{jialiu@pku.edu.cn}
\affiliation{School of Physics and State Key Laboratory of Nuclear Physics and Technology, Peking University, Beijing 100871, China.}
\affiliation{
Center for High Energy Physics, Peking University, Beijing 100871, China
}

\author{Masanori Tanaka}
\email{tanaka@pku.edu.cn}
\affiliation{
Center for High Energy Physics, Peking University, Beijing 100871, China
}

\author{Xiao-Ping Wang}
\email{hcwangxiaoping@buaa.edu.cn}
\affiliation{School of Physics, Beihang University, Beijing 100083, China}

\author{Jing-Jun Zhang}
\email{zhang\_jingjun@stu.pku.edu.cn}
\affiliation{School of Physics and State Key Laboratory of Nuclear Physics and Technology, Peking University, Beijing 100871, China.}

\author{Zifan Zheng}
\email{zifanzheng@pku.edu.cn}
\affiliation{School of Physics and State Key Laboratory of Nuclear Physics and Technology, Peking University, Beijing 100871, China.}

\date{\today}

\preprint{$\begin{gathered}\includegraphics[width=0.05\textwidth]{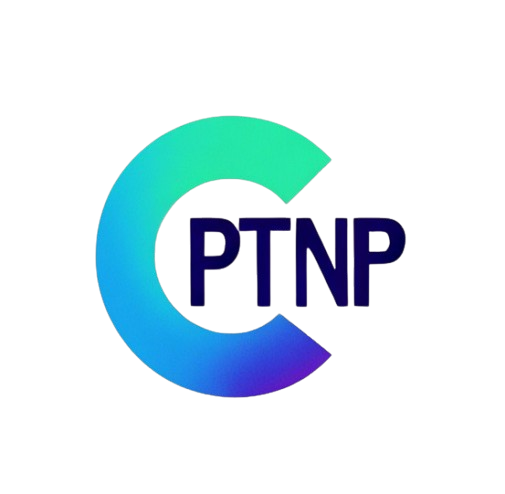}\end{gathered}$\, CPTNP-2025-010}

\begin{abstract}
We investigate the connection between the entanglement entropy in scattering processes and the dynamics of electroweak phase transitions. 
Recent work has shown that the scattering entanglement entropy can provide new insight into Standard Model parameters. 
In this study, we propose that the maximum of the entanglement entropy in scattering amplitudes may serve as a diagnostic for first-order electroweak phase transitions in the early universe. 
We analyze a simplified extension of the Standard Model consisting of the Higgs boson $h$ coupled to $O(N)$ real singlet scalars $S$ via the Higgs portal coupling $\lambda_{hS}$. 
By explicitly calculating the maximum entanglement entropy, we demonstrate that it grows with increasing $\lambda_{hS}$, and that both first-order and strong first-order electroweak phase transitions are favored in regions of parameter space with large maximum entropy. 
Our results suggest that entanglement-based observables may encode meaningful information about the underlying dynamics of electroweak symmetry breaking and provide a novel perspective on phase transition phenomena.

\end{abstract}

\maketitle
\newpage

\tableofcontents

\section{Introduction}

The significance of quantum entanglement in quantum field theories has gained increasing recognition in recent years. 
One of the key quantities used to quantify the strength of quantum entanglement is the entanglement entropy, which has found applications across various domains. 
For instance, in the context of black hole thermodynamics, the entanglement entropy plays a central role~\cite{Bekenstein:1973ur, Hawking:1975vcx}. 
It has been suggested that the entropy of black holes can be interpreted as entanglement entropy, with the degrees of freedom inside the black hole being traced out~\cite{Bombelli:1986rw}.
Beyond black holes, the role of entanglement entropy in scattering processes has also attracted considerable attention~\cite{Seki:2014cgq, Peschanski:2016hgk, Cervera-Lierta:2017tdt, Peschanski:2019yah, Beane:2018oxh, Low:2021ufv, Beane:2021zvo, Bai:2022hfv, Fedida:2022izl, Miller:2023ujx, Liu:2022grf, Liu:2023bnr, Fedida:2024dwc, Blasone:2024dud, Hu:2024hex, Chang:2024wrx, Kowalska:2024kbs, Carena:2023vjc, Aoude:2024xpx, Low:2024hvn, Thaler:2024anb, Blasone:2024jzv, McGinnis:2025brt, Carena:2025wyh, Blasone:2025tor}. 
One of the most intriguing findings is that minimizing the variation of the entanglement entropy during scattering processes can lead to the prediction of symmetries in new physics models~\cite{Beane:2018oxh, Low:2021ufv, Beane:2021zvo, Bai:2022hfv, Miller:2023ujx, Liu:2022grf, Liu:2023bnr, Hu:2024hex, Chang:2024wrx, Kowalska:2024kbs, Carena:2023vjc, McGinnis:2025brt, Carena:2025wyh}. 
Moreover, these insights can potentially explain experimentally measured quantities. 
For example, the requirement of maximal entanglement entropy between the helicity states of leptons in QED scattering processes has been shown to predict the experimentally observed Weinberg angle~\cite{Cervera-Lierta:2017tdt}. 
Similarly, minimizing entanglement entropy between flavor states in scattering processes of left-handed quarks and leptons can simultaneously explain the structure of the Cabibbo–Kobayashi–Maskawa matrix and the Pontecorvo-Maki-Nakagawa-Sakata matrix~\cite{Thaler:2024anb}. 
These findings suggest that entanglement entropy may provide valuable insights into the particle physics.

The entanglement entropy has been a key focus in condensed matter physics (e.g. see Ref.~\cite{Laflorencie:2015eck}), where it serves as a powerful diagnostic tool for quantum phase transitions, especially those involving topological orders~\cite{Kitaev:2005dm, Levin:2006zz}. 
These studies highlight the potential of entanglement entropy as an order parameter in phase transitions.

Moreover, it has been shown that holographic entanglement entropy effectively describes the confinement/deconfinement phase transitions in non-Abelian gauge theories.
In the context of the 5-dimensional Anti-de Sitter (AdS) soliton solution, the corresponding dual gauge theory is the $\mathcal{N} = 4$ supersymmetric Yang–Mills theory on $R^{1,2} \times S^{1}$, due to gauge/gravity duality~\cite{Witten:1998zw}. 
Using this duality, the entanglement entropy in the gauge theory can be related to geodesics in the gravity side~\cite{Ryu:2006bv}. 
The holographic entanglement entropy between the inside and outside of a domain with width $\ell$ has been calculated in the AdS background~\cite{Klebanov:2007ws, Nishioka:2006gr,Barbosa:2024pyn}. 
The derivative of entanglement entropy with respect to $\ell$ is discrete at a critical value $\ell^*$, with no variation for $\ell > \ell^*$.
This critical value corresponds to the phase transition temperature for confinement/deconfinement. 
These results have been confirmed numerically in lattice gauge theories, such as the $\rm SU(3)$ gauge theory~\cite{Nakagawa:2009jk, Itou:2015cyu}.
Additionally, the entanglement entropy between valence and non-valence spin systems has been identified as an order parameter for chiral symmetry breaking in nucleons~\cite{Beane:2019loz}. 
These findings inspire our investigation of the connection between the entanglement entropy and phase transitions in the early universe.

On the other hand, the phase transitions in the early universe are thought to be crucial in explaining phenomena beyond the Standard Model (SM), such as the baryon asymmetry of the universe (BAU)~\cite{Planck:2018vyg}. 
A mechanism satisfying Sakharov's conditions is required to generate the observed BAU~\cite{Sakharov:1967dj}. 
Among the potential mechanisms, the electroweak (EW) baryogenesis scenario is particularly promising~\cite{Kuzmin:1985mm}. 
This scenario requires a first-order electroweak phase transition (FOEWPT) to satisfy the Sakharov's conditions for departure from thermal equilibrium. 
To avoid washout by EW sphaleron processes, the electroweak phase transition (EWPT) should meet the following condition~\cite{Kuzmin:1985mm}:
\begin{align}
\label{eq:vcTc1}
\frac{v_{C}}{T_{C}} > \zeta_{\rm sph}(T_{C}) \,, 
\end{align}
where $T_{C}$ is the critical temperature, and $v_{C}$ is the vacuum expectation value (VEV) for the SM Higgs field at $T = T_{C}$. 
In particular, the FOEWPT that satisfies the condition \eqref{eq:vcTc1} is called the strongly first-order electroweak phase transition (sFOEWPT). 
While $\zeta_{\rm sph}(T)$ depends slightly on new physics models, a representative value is often taken as $\zeta_{\rm sph}(T_{C}) \simeq 1$~\cite{Ahriche:2007jp, Funakubo:2009eg, Fuyuto:2014yia, Ahriche:2014jna, Kanemura:2020yyr, Kanemura:2022ozv}. 
Lattice simulations within the SM framework have shown that the EWPT in the SM is a crossover~\cite{Kajantie:1996mn,Csikor:1998eu,Aoki:1999fi,DOnofrio:2014rug,DOnofrio:2015gop}, meaning the SM cannot satisfy the condition in Eq.\,\eqref{eq:vcTc1}. 
This result indicates the need for new physics models to realize EW baryogenesis and sFOEWPT. 
The feasibility of EW baryogenesis in extended Higgs models has been extensively studied~\cite{Turok:1990in, Turok:1990zg, Cline:1995dg, Carena:1997gx, Cline:1997vk, Cline:1998hy, Cline:2000kb, Cline:2000nw, Carena:2000id, Kainulainen:2002th, Carena:2002ss, Cirigliano:2006wh, Fromme:2006cm, Chung:2009qs, Li:2008ez, Cline:2011mm, Liu:2011jh, Cline:2012hg, Chiang:2016vgf, Jiang:2015cwa, Guo:2016ixx, Vaskonen:2016yiu, Fuyuto:2017ewj, Grzadkowski:2018nbc, Modak:2018csw, Cline:2021iff, Kainulainen:2021oqs, Xie:2020wzn, Enomoto:2021dkl, Enomoto:2022rrl, Aoki:2022bkg, Aoki:2023xnn, Basler:2021kgq, Kanemura:2023juv, Aiko:2025tbk}.

In this paper, we investigate the connection between the entanglement entropy of scattering processes and the dynamics of phase transitions in the early universe.
Since the entanglement entropy is closely linked to symmetries in new physics models, it naturally suggests a relationship with the dynamics of symmetry breaking and phase transitions. 
Thus, we study a simplified model consisting of the SM Higgs doublet field $H$, coupled to $N$ real scalar fields $S_{\alpha}$ ($\alpha = 1, \dots, N$) possessing a global $O(N)$ symmetry. 
We assume that $H$ induces the symmetry breaking, while the other fields do not.
By setting $N=4$, the model becomes analogous to two-Higgs-doublet models (2HDMs). We analyze the entanglement entropy in scattering processes between the scalar fields in this model and demonstrate that the sFOEWPT tends to favor larger entanglement entropy. 
Our findings suggest that the entanglement entropy may provide a plausible explanation for the occurrence of sFOEWPTs in the early universe.
Furthermore, we explore the possibility that the entanglement entropy could serve as an order parameter for the EWPT.

The structure of this paper is as follows.
In Section\,\ref{sec:ONsinglet}, we introduce the $O(N)$ singlet models. 
In Section\,\ref{sec:fopt}, the thermal effective potential in the $O(N)$ singlet model is defined. 
In Section\,\ref{sec:entanglement}, the definition of entanglement entropy in the scalar scattering process is defined. 
Our main numerical results are shown in Section\,\ref{sec:numericalresults}. 
Then, the discussion and conclusion are summarized in Section\,\ref{sec:discussion} and Section\,\ref{sec:conclusion}, respectively.

\section{$O(N)$ singlet models \label{sec:ONsinglet}}

In this section, we introduce $O(N)$ singlet models, which include two types of scalar fields: the SM Higgs doublet field $H$ and new singlet fields $\vec{S} = (S_1, \cdots, S_N)^{\rm T}$. 
The field $H$ is responsible for inducing the spontaneous EW symmetry breaking, while $\vec{S}$, although it does not acquire a VEV, can influence EWPT dynamics through radiative corrections. 
The model assumes a global $O(N)$ symmetry for the scalar field $\vec{S}$, which leads to the following scalar potential:
\begin{align}
\begin{aligned}
V_{0}(H, \vec{S}) 
= - \mu_{h}^2 |H|^2 + \frac{\mu_{S}^2}{2} (\vec{S} \cdot \vec{S}) + \lambda_{h} |H|^4 + \frac{\lambda_{S}}{4} (\vec{S} \cdot \vec{S})^2 + \frac{\lambda_{hS}}{2} |H|^2 (\vec{S} \cdot \vec{S}) \,.
\end{aligned}
\end{align}
To avoid the symmetry breaking along the $\vec{S}$ direction, the condition $\mu_{S}^2 \geq 0$ is imposed in our study. 

To realize the spontaneous symmetry breaking along the $H$ direction, the following conditions are imposed: 
\begin{align}
\label{eq:stat-cond}
\begin{aligned}
&\left. \frac{\partial V}{\partial \phi} \right|_{\braket{\phi} = v_{h}, \, \braket{\vec{S}} = \vec{0}} = 0 \,, \\ 
&\left. \frac{\partial^2 V}{\partial \phi^2} \right|_{\braket{\phi} = v_{h}, \, \braket{\vec{S}} = \vec{0}} = m_{h}^2 \,,
\end{aligned}
\end{align}
where $H = (0, \phi/\sqrt{2})^{\rm T}$, $m_{h} = 125\,{\rm GeV}$ and $v_{h} =246\,{\rm GeV}$. 
The masses of the fields $h = \phi - v_{h}$ and $\vec{S}$ at the vacuum are given by 
\begin{align}
\label{eq:Smass}
m_{h}^2
&= - \mu_{h}^2 + 3\lambda_{h} v_{h}^2 = 2 \lambda_{h} v_{h}^2\,, \\
m_{S}^2 &= \mu_{S}^2 + \frac{\lambda_{hS}}{2} v_{h}^2 \,. 
\end{align}
In the regime where $m_S \simeq \mu_S$, the field $\vec{S}$ exhibits a decoupling behavior, meaning its effects on low-energy observables become negligible as $m_S$ increases. This property, known as decoupling, has been well studied in the literature~\cite{Appelquist:1974tg}. However, when $m_S^2 \simeq \lambda_{hS} v_h^2/2 \gg \mu_S^2$, the effects of $\vec{S}$ on low-energy observables can remain significant due to large radiative corrections, even if $m_S$ is large. This non-decoupling behavior has been found to be crucial for realizing a sFOEWPT~\cite{Kanemura:2004ch}. The differences between decoupling and non-decoupling cases have been extensively studied using effective field theoretical approaches~\cite{Kanemura:2021fvp, Banta:2022rwg, Kanemura:2022txx, Florentino:2024kkf}.

As will be discussed later, the sFOEWPT typically requires a large $\lambda_{hS}$. As a theoretical constraint on such large scalar field coupling constants, we incorporate the perturbative unitarity bound~\cite{Lee:1977eg}. For the $O(N)$ singlet model, this bound is expressed as~\cite{Hashino:2016rvx}:
\begin{align}
\label{eq:unitarity}
\begin{aligned}
6 \lambda_{h} &+ (N+2) \lambda_{S} + \sqrt{ \left\{ 6 \lambda_{h} - (N+2) \lambda_{S} \right\}^2 + 4 N \lambda_{hS}^2} < 16 \pi \,.
\end{aligned}
\end{align}
This unitarity bound is applied as a constraint in our numerical analysis.

Before we move into the next section, we note that our model includes scalar singlet fields $\vec{S}$ with the global $O(N)$ symmetry. Because of this symmetry, $\vec{S}$ can naturally be considered a dark matter (DM) candidate. As a result, the model is subject to limits from current DM direct detection experiments, such as CDEX~\cite{CDEX:2019hzn}, XENONnT~\cite{XENON:2023cxc}, PandaX-4T~\cite{PandaX:2024qfu}, and LZ~\cite{LZ:2024zvo}. However, our numerical results show that the parameter region we focus on is already excluded by these experiments. One way to avoid these DM constraints is to assume that the $O(N)$ symmetry is not exact or slightly broken. For example, it could be broken by additional interactions like Yukawa couplings. In this case, the singlet field $\vec{S}$ would no longer be stable and cannot serve as a DM particle. Thus, direct detection limits would not apply. However, the above modifications will not change our conclusions about the scalar potential and its role in the EWPT dynamics.

\section{Effective potential at finite temperatures \label{sec:fopt}}

In this section, we study the effective potential at finite temperatures. The total effective potential is given by:
\begin{align}
V_{\rm eff}(\phi, T) 
= V_{0}(\phi, \vec{S} = \vec{0}) + V_{\rm CW}(\phi) + V_{T}(\phi, T) \,.
\end{align}
The second term, $V_{\rm CW}(\phi)$, accounts for the radiative corrections at zero temperature and is given by~\cite{Coleman:1973jx}:
\begin{align}
V_{\rm CW}(\phi) = \sum_{\alpha} \frac{n_{\alpha}}{64\pi^2} m_{\alpha}^4(\phi) \left[ \ln \left( \frac{m_{\alpha}^2(\phi)}{\mu^2} \right) - \frac{3}{2} \right] \,,
\end{align}
where $m_{\alpha}^2(\phi)$ representing the field-dependent masses for particle $\alpha$. 
In our analyses, we take into account the contributions from weak gauge bosons $(W, Z)$ and the top quark $(t)$ in addition to the scalar fields: $\alpha = \{ h, \vec{S}, W, Z, t \}$ and $n_{\alpha} = \{1, N, 6, 3, -12 \}$. 
The renormalization scale $\mu$ is chosen so that the tree-level value of the mass for the field $h$ remains unchanged by higher-order corrections~\cite{Anderson:1991zb}. 
The third term, $V_T(\phi, T)$, describes the thermal corrections and is given by~\cite{Dolan:1973qd}:
\begin{align}
V_{T}(\phi, T)
= \frac{T^4}{2 \pi^2} \sum_{\alpha} n_{\alpha} J_{\alpha}\left(\frac{m_{\alpha}^2(\phi)}{T^2}\right) \,,
\end{align}
where the function $J_{\alpha}(a^2)$ is defined as:
\begin{align}
J_{\alpha}\left(a^2\right)
= \int_{0}^{\infty} dk k^2 \ln \left[ 
1 - {\rm sign}(n_{\alpha}) e^{-\sqrt{ k^2 + a^2 }}
\right] \,,
\end{align}
with ${\rm sign}(n_{\alpha})$ being $+1$ or $-1$ depending on whether $n_{\alpha}$ is positive or negative.

At high temperatures, the infrared divergence in the zero-mode propagator for bosonic fields must be properly handled~\cite{Carrington:1991hz, Parwani:1991gq,Arnold:1992rz}. 
We use the method developed by Parwani~\cite{Parwani:1991gq}, which replaces the field-dependent masses in $V_{\rm CW}(\phi)$ and $V_T(\phi, T)$ with their thermal masses. 
The thermal masses for $\phi$ and $\vec{S}$ are~\cite{Espinosa:2008kw}:
\begin{align}
\begin{aligned}
m_{h}^2(\phi, T)
&= - \mu_{h}^2 + 3\lambda_{h} \phi^2 + T^2 \left( \frac{\lambda_{h}}{2} + \frac{g_{1}^2}{16} + \frac{3g_{2}^2}{16} + \frac{y_{t}^2}{4}+ \frac{\lambda_{hS}}{24} N \right) \,, \label{eq:thermalmass}
 \\
m_{S}^2(\phi, T)
&= \mu_{S}^2 + \frac{\lambda_{hS}}{2} \phi^2 + \frac{T^2}{12} \left[ \lambda_{S} (N+2) + 2\lambda_{hS} \right] \,,
\end{aligned}
\end{align}
where $g_{1}, g_{2}$ and $y_{t}$ are the $U(1)_{Y}$ gauge, $SU(2)_{L}$ gauge and top quark Yukawa couplings, respectively. 
For the field content in the SM $(W, Z, \,t)$, we employ the formalism discussed in Ref.\,\cite{Carrington:1991hz}. 

The $O(N)$ singlet model is a useful approximation to describe EWPTs in various new physics models with extended Higgs sectors~\cite{Kakizaki:2015wua}.
For example, in 2HDMs, the strength of the FOEWPT in the high-temperature and non-decoupling limit is given by~\cite{Kanemura:2004ch}:
\begin{align}
\label{eq:vcTc_2HDM}
\frac{v_{C}}{T_{C}} \simeq \frac{4}{3 \pi v_{h} m_{h}^2} \left( m_{H}^3 + m_{A}^3 + 2 m_{H^{\pm}}^3 \right) \,,
\end{align}
where we neglect contributions from the $W$ and $Z$ bosons for simplicity. The parameters $m_\Phi^2 ~ (\Phi = H, A, H^\pm)$ are the physical masses of the heavy Higgs bosons in the 2HDM. For the $O(N)$ singlet model, we find a similar expression for the FOEWPT: 
\begin{align}
\label{eq:vcTc_ON}
\frac{v_{C}}{T_{C}} \simeq \frac{4 N m_{S}^3 }{3 \pi v_{h} m_{h}^2}  \,.
\end{align}
This shows that the $O(N)$ singlet model can approximately describe the dynamics of the EWPT in the 2HDM when we set $N = 4$ and $m_S = m_\Phi~ (\Phi = H, A, H^\pm)$. 
For $N = 1$ or $N = 2$, the $O(N)$ singlet model corresponds to the SM extended with a real or complex singlet scalar field.
Thus, we use the $O(N)$ singlet model to derive a general relationship between the entanglement entropy and the EWPT. 

For the criterion of the FOEWPT, we utilize the lattice simulation results because it is difficult to determine the order of EWPT in the perturbation theory. 
According to the lattice simulation for the EWPT~\cite{Kajantie:1996mn,Csikor:1998eu,Aoki:1999fi}, the EWPT can be first order if $m_{h} \lesssim 80\,{\rm GeV}$. 
We have numerically obtained $\left. v_{C}/T_{C} \right|_{\rm SM} \simeq 0.24$ in the SM with $m_{h} = 80\,{\rm GeV}$ without the high temperature limit in the perturbation theory. 
In our analyses, we conclude that the EWPT can be first order when the following condition is satisfied
\begin{align}
\label{eq:fopt}
\frac{v_{C}}{T_{C}} \geq \left. \frac{v_{C}}{T_{C}} \right|_{\rm SM} \simeq 0.24 \,.
\end{align}
Obviously, this criterion is rather rough.
However, since we do not have the result of the lattice simulation on the EWPT in the $O(N)$ singlet model, we employ this criterion as an approximate criterion.

\section{Definitions of measures for the entanglement \label{sec:entanglement}}

In this section, we introduce the concept of entanglement entropy for scattering processes, following the formalism developed in Refs.~\cite{Kowalska:2024kbs, Low:2024hvn}. We consider a two-particle system and define the total Hilbert space for particles $a$ and $b$ as:
\begin{align}
\mathcal{H} = \mathcal{H}^{a}_{\rm kin} \otimes \mathcal{H}^{a}_{f} \otimes \mathcal{H}^{b}_{\rm kin} \otimes \mathcal{H}^{b}_{f} \,, 
\end{align}
where $\mathcal{H}_{\rm kin}$ and $\mathcal{H}_{f}$ represent the momentum and intrinsic flavor Hilbert spaces for each particle, respectively. 
We assume that the dimension of $\mathcal{H}_{f}^{a, b}$ takes $G$. 

The general state $\ket{\Psi}$ for the bi-particle system can be expressed by~\cite{Aoude:2024xpx, Kowalska:2024kbs, Low:2024hvn}
\begin{align}
    \ket{\Psi}
    = \sum_{\alpha, \beta = 1}^{G} c_{\alpha \beta} \ket{\psi_{a}} \otimes \ket{\alpha}_{a} \otimes \ket{\psi_{b}} \otimes \ket{\beta}_{b} \,,
\end{align}
where $\ket{\psi_{a/b}}$ corresponds to the kinetic states for each particle. 
The states $\ket{\alpha}_{a}$ and $\ket{\beta}_{b}$ represent the flavor states of the particles $a$ and $b$. 
The coefficient $c_{\alpha \beta}$ is determined by the normalization condition $\braket{\Psi|\Psi} = 1$. 

Once the state $\ket{\Psi}$ is obtained, we can compute the entanglement entropy for the bi-partite system $A$ and $B$. The Hilbert spaces corresponding to these subsystems are denoted as $\mathcal{H}^{A}$ and $\mathcal{H}^{B}$, respectively. In this work, we focus on the linear entropy as a measure of entanglement. Specifically, the entanglement entropy for subsystem $A$ is defined as:
\begin{align}
\label{eq:linearEE}
E(\rho) = 1 - {\rm tr}_{A}(\rho_{A}^2)  \,,
\end{align}
where $\rho_{A}$ is the reduced density matrix of subsystem $A$.
The reduced density matrix is obtained by tracing out the degrees of freedom of subsystem $B$:
\begin{align}
\rho_{A} \equiv {\rm tr}_{B}^{} (\rho) \,.
\end{align} 

We here consider the entanglement entropy in the following $2\to2$ scattering processes
\begin{align}
\label{eq:process}
S_{\alpha} (p_{a}) S_{\beta} (p_{b}) \to S_{\gamma} (p_{c}) S_{\delta} (p_{d}) \,.
\end{align}
In this setup, we can take $G = N$, and we will assume the center-of-mass frame for simplicity.

To obtain the reduced density matrix, we must trace out the information associated with the state defined in $\mathcal{H}^{B}$. 
In this paper, we focus on the entanglement within the bi-partite system 
\begin{align}
\mathcal{H}^{A} \equiv \mathcal{H}^{a}_{\rm kin} \otimes \mathcal{H}^{a}_f \,, \quad \mathcal{H}^{B} \equiv \mathcal{H}^{b}_{\rm kin} \otimes \mathcal{H}^{b}_f \,,     
\end{align}
corresponding to the entanglement entropy between particles $a$ and $b$. 
While alternative bipartite systems, such as $\mathcal{H}^{a}_f \otimes \mathcal{H}^{b}_f$ and $\mathcal{H}^{a}_{\rm kin} \otimes \mathcal{H}^{b}_{\rm kin}$~\cite{Low:2024hvn}, can be considered, numerical analyses confirm that our conclusions are not significantly affected by the choice of entanglement definition.
The bi-partite system we consider is particularly well-suited for our analysis because the scattering process in Eq.~\eqref{eq:process} may involve flavor-identical particles in the initial or final states. 
In such cases, the flavor-identical particles can be distinguished by their kinetic states, which is crucial for ensuring that the entanglement entropy remains well-defined. 
Therefore, we focus on the entanglement entropy defined in the bipartite system $\mathcal{H}^{A}$ and $\mathcal{H}^{B}$ as the representative result.

To compute the density matrix, we first specify the initial state in the scattering process. The general initial state is given by~\cite{Kowalska:2024kbs, Low:2024hvn}
\begin{align}
\label{eq:isIJ}
\begin{aligned}
\ket{\rm in}_{\alpha \beta} = \int \int & \frac{d^3 p_{1}}{(2\pi)^3 \sqrt{2E_{p_{1}}}} \frac{d^3 p_{2}}{(2\pi)^3 \sqrt{2E_{p_{2}}}} \psi_{a}(\mathbf{p_{1}}) \psi_{b}(\mathbf{p}_{2})
\ket{ \mathbf{p}_{1}, \alpha} \ket{ \mathbf{p}_{2}, \beta} \,. 
\end{aligned}
\end{align}
with 
\begin{align}
&E_{p_{1},p_{2}} = \sqrt{|\mathbf{p}_{1,2}|^2 + m_{\alpha,\, \beta}^2} \,,  \\
&\braket{ \mathbf{p}_{1}, \alpha | \mathbf{p}_{2}, \beta} = (2 \pi)^3 2 E_{p_{1}} \delta_{\alpha \beta} \delta^{3}( \mathbf{p}_{1} - \mathbf{p}_{2}) \,,
\end{align}
where $m_{\alpha}$ is the mass of $S$ field with the flavor $\alpha$. In the plane wave limit, the wave functions $\psi_{a/b}(\mathbf{p})$ are given by:
\begin{align}
\psi_{a, \, b}(\mathbf{p}) = \sqrt{\frac{(2\pi)^3}{\delta^3(0)}} \delta^{3}(\mathbf{p}_{a, \, b} - \mathbf{p}) \,, 
\end{align}
where we assumed that the particles $a$ and $b$ have the momenta $\mathbf{p}_{a}$ and $\mathbf{p}_{b}$, respectively. 
The prefactor of the delta function are required to satisfy the normalization condition $|\ket{ {\rm in}}_{\alpha \beta}|^2 = 1$. 

We note that the initial flavor state with $\alpha=\beta$ in Eq.\,\eqref{eq:isIJ} is not distinguishable. 
However, the particles $a$ and $b$ with the same flavor state can be kinematically distinguishable if the wave packet formalism is used~\cite{Kowalska:2024kbs, Low:2024hvn}. 
Importantly, as demonstrated in Refs.\,\cite{Kowalska:2024kbs, Low:2024hvn}, the predictions for the entanglement entropy remain unchanged even in the plane-wave limit.
Therefore, we adopt the wave packet formalism incorporating the plane-wave limit to investigate the entanglement properties. 
In the plane-wave limit, the initial state is given by~\cite{Kowalska:2024kbs}:
\begin{align}
\ket{\rm in}_{\alpha \beta}
= \frac{1}{X} \ket{ \mathbf{p}_{a}, \alpha} \ket{ \mathbf{p}_{b}, \beta} \,, 
\end{align}
with $X = (2\pi)^3 \delta^{3}(0) \sqrt{4 E_{p_{a}} E_{p_{b}}}$. 

Next, we define the final state, which can be experimentally determined. 
The final state $\ket{\rm out}_{\gamma \delta}$ is obtained by acting with the scattering operator $\mathcal{S}$ and a projection operator on the initial state. 
This process is described by~\cite{Aoude:2024xpx, Kowalska:2024kbs, Low:2024hvn}:
\begin{align}
\begin{aligned}
\ket{ {\rm out} }_{\gamma \delta} 
&= \iint d^{3} \tilde{p}_{c} d^{3} \tilde{p}_{d} (\mathcal{S}_{f})_{\alpha \beta \gamma \delta}^{abcd} \ket{\mathbf{p}_{c}, \gamma ; \mathbf{p}_{d}, \delta}\,, 
\end{aligned}
\end{align}
with 
\begin{align}
\int d^{3} \tilde{p}_{c} \equiv \int \frac{d^3 p_{c}}{(2\pi)^3 2 E_{p_{c}}} \,.
\end{align}
The operator $\mathcal{S}$ is the conventional S-matrix operator. 
The indices $\gamma$ and $\delta$ represent the flavors in the final state, which is part of the post-scattering state $\ket{{\rm out}} = \mathcal{S} \ket{{\rm in}}_{\alpha \beta}$.
The matrix element of the scattering operator $\mathcal{S}_{f}$ operator in the basis $\left\{ \bra{\mathbf{p}_{c}, \gamma ; \mathbf{p}_{d}, \delta}, \, \ket{\mathbf{p}_{a}, \alpha ; \mathbf{p}_{b}, \beta} \right\}$ is given by 
\begin{align}
(\mathcal{S}_{f})_{\alpha \beta \gamma \delta}^{abcd} &\equiv \braket{\mathbf{p}_{c}, \gamma ; \mathbf{p}_{d}, \delta | \mathcal{S}_{f} | \mathbf{p}_{a}, \alpha ; \mathbf{p}_{b}, \beta} \nonumber \\
&= (2\pi)^6 4 E_{p_{a}} E_{p_{b}} \delta_{\alpha \beta \gamma \delta}^{abcd} + (2\pi)^4 \delta^{4}(p_{a} + p_{b} - p_{c} - p_{d})  i \mathcal{M}_{\alpha \beta \gamma \delta}(p_{a}, p_{b} \to p_{c}, p_{d})  \,,
\end{align}
with 
\begin{align}
\delta_{\alpha \beta \gamma \delta}^{abcd} \equiv
\delta_{\alpha \gamma} \delta_{\beta \delta} \delta^{3}( \mathbf{p}_{a} - \mathbf{p}_{c}) \delta^{3}( \mathbf{p}_{b} - \mathbf{p}_{d}) \,.
\end{align}
As the operator $\mathcal{S}_{f}$ is not a unitary operator, the normalization must be applied when evaluating the density matrix for the final state. 
Then, the final state can be written as
\begin{align}
\label{eq:finalstate}
&\ket{\rm out}_{\gamma \delta} = \frac{1}{X} \left[ 
\delta_{\alpha \gamma} \delta_{\beta \delta} \ket{\mathbf{p}_{a}, \gamma ; \mathbf{p}_{b}, \delta} \right. \nonumber \\ &\left.
\quad + \iint d^3 \tilde{p}_{c} d^3 \tilde{p}_{d}
(2\pi)^4 \delta^{4}(p_{a} + p_{b} - p_{c} - p_{d}) i \mathcal{M}_{\alpha \beta \gamma \delta}
(p_{a}, p_{b} \to p_{c}, p_{d}) \ket{\mathbf{p}_{c}, \gamma ; \mathbf{p}_{d}, \delta}
\right] \,. 
\end{align}
Our definition of the final state differs from that in Ref.\,\cite{Thaler:2024anb}. In Ref.\,\cite{Thaler:2024anb}, the final state is defined as one in which the kinetic properties are fixed by measurements. Thus, the kinetic state does not need to be traced out. In contrast, since we do not use a post-measurement state, the kinetic degrees of freedom must be traced out at the amplitude level~\cite{Kowalska:2024kbs}.

In our approach, the final state, as given in Eq.\,\eqref{eq:finalstate}, leads to the following density matrix:
\begin{align}
\label{eq:densitymatrix}
(\rho_{f})_{\gamma \delta \eta \zeta}
= \frac{ \ket{\rm out}_{\gamma \delta} \bra{\rm out}_{\eta \zeta} }{{\rm tr}_{AB}^{}\left( \ket{\rm out}_{\gamma \delta} \bra{\rm out}_{\eta \zeta} \right)} \,. 
\end{align}
Here, ${\rm tr}_{AB}$ denotes the trace over all degrees of freedom associated with the subsystems $A$ and $B$ such as the flavor states. The denominator ensures that the density matrix satisfies the normalization condition ${\rm tr}_{AB}^{}[\rho_{f}] = 1$. Consequently, the reduced density matrix $\rho_{f}$ for the subsystem $A$ in the final state is defined by:
\begin{align}
\begin{aligned}
(\rho^{A}_{\alpha \beta})_{\gamma \eta} = {\rm tr}_{B}(\rho_{f}) 
&= \sum_{\delta, \zeta, \bar{\delta}} \int d^3 \tilde{p}_{2} \braket{\mathbf{p}_{2}, \bar{\delta}|_{b}^{} (\rho_{f})_{\gamma \delta \eta \zeta} | \mathbf{p}_{2}, \bar{\delta}}_{b}^{} \\
&= \frac{ 1 }{ {\rm tr}_{AB}^{} \left[ \ket{\rm out}_{\gamma \delta} \bra{\rm out}_{\eta \zeta}  \right]}
\sum_{\delta, \zeta, \bar{\delta}} \int d^3 \tilde{p}_{2} \braket{\mathbf{p}_{2}, \bar{\delta}|_{b}^{} {\rm out}}_{\gamma \delta} \braket{ {\rm out}|_{\eta \zeta} \mathbf{p}_{2}, \bar{\delta} }_{b}^{} \,.
\end{aligned}
\end{align}
where the label $b$ indicates that the trace is performed over the degrees of freedom of particle $b$. The indices $\alpha$ and $\beta$ represent the flavor of the initial state. 
The summation over the flavor indices $\delta$ and $\zeta$, along with $\bar{\delta}$, is necessary because these indices correspond to the flavor of the particle $b$.
For the integral part, we obtain 
\begin{align}
\label{eq:rhoAinte}
&\sum_{\delta, \zeta, \bar{\delta}}  \int d^3 \tilde{p}_{2} \braket{\mathbf{p}_{2}, \bar{\delta}|_{b}^{} {\rm out}}_{\gamma \delta} \braket{ {\rm out}|_{\eta \zeta} \mathbf{p}_{2}, \bar{\delta} }_{b}^{} \nonumber \\
&\equiv \frac{1}{X^2} \left[ B_{\gamma \eta } \ket{\mathbf{p}_{a}, \gamma} \bra{\mathbf{p}_{a}, \eta} 
 + \sum_{\delta}
\int \frac{d^3 p_{c}}{(2\pi)^3} \frac{(2\pi)^2}{8 (E_{p_{c}})^3} \mathcal{M}_{\alpha \beta \gamma \delta} \mathcal{M}_{\alpha \beta \eta \delta}^{*} [\delta(E_{p_{a}} + E_{p_{b}} - 2 E_{p_{c}} )]^2 \ket{\mathbf{p}_{c}, \gamma} \bra{\mathbf{p}_{c}, \eta} \right] \,, 
\end{align}
where
\begin{align}
\begin{aligned}
B_{\gamma \eta} 
=& 
(2\pi)^3 2 E_{p_{b}} \delta^{3}(0) \delta_{\alpha \gamma} \delta_{\alpha \eta} \\
&+ \frac{ 2\pi }{2 E_{p_{a}}}  \delta(0) \left[ i \delta_{\alpha \eta} \mathcal{M}_{\alpha \beta \gamma \beta} (p_{a}, p_{b} \to p_{a}, p_{b}) - i \delta_{\alpha \gamma} \mathcal{M}_{\alpha \beta \eta \beta}^{*} (p_{a}, p_{b} \to p_{a}, p_{b}) \right] \,.
\end{aligned}
\end{align}
Therefore, the trace of squared reduced density matrix can be given by 
\begin{align}
\begin{aligned}
&{\rm tr}_{A}  \left[ (\rho_{\alpha \beta}^{A})^2 \right]= \frac{ 1 }{ {\rm tr}_{AB}\left[ \ket{\rm out}_{\gamma \delta} \bra{\rm out}_{\eta \zeta}  \right]^2}  \frac{1}{X^4} \\
&\quad \times \sum_{\gamma,\, \eta}\left[
B_{\gamma \eta} B_{\eta \gamma}^{*} 4 E_{p_{a}}^2 \left\{ (2\pi)^3 \delta^{3}(0) \right\}^2 
+ \frac{B_{\gamma \eta}}{2E_{p_{a}}} (2\pi)^5 \delta^{3}(0) \left\{ \delta (0) \right\}^2 \sum_{\delta'} \mathcal{M}_{\alpha \beta \eta \delta'}^{*} \mathcal{M}_{\alpha \beta \gamma \delta'} \right. \\ & \left. 
\quad + \frac{B_{\eta \gamma}^{*}}{2E_{p_{a}}} (2\pi)^5 \delta^{3}(0) \left\{ \delta(0) \right\}^2 \sum_{\delta} \mathcal{M}_{\alpha \beta \eta \delta} \mathcal{M}_{\alpha \beta \eta \delta}^{*} \right.  \\ & \left. 
\quad + (2\pi)^4 \left\{ \delta (0) \right\}^4 \delta^{3}(0) 
\sum_{\delta, \delta'} \int \frac{d^3 p_{c1} }{16 E_{p_{c1}}^4 } \mathcal{M}_{\alpha \beta \gamma \delta} \mathcal{M}_{\alpha \beta \eta \delta}^{*} \mathcal{M}_{\alpha \beta \eta \delta'}^{*} \mathcal{M}_{\alpha \beta \gamma \delta'} 
\right] \,,
\end{aligned}
\end{align}
where $\delta(0)$ and $\delta^3(0)$ come from the delta function for the energy and momentum conservation, respectively. 
By utilizing $\Delta \equiv (2\pi)^4 \delta(0) \delta^3(0)/X^2$, we obtain
\begin{align}
\label{eq:trrhoA}
\begin{aligned}
&{\rm tr}_{A}  \left[ (\rho_{\alpha \beta}^{A})^2 \right] =
\frac{ 1 }{ {\rm tr}_{AB}\left[ \ket{\rm out}_{\gamma \delta} \bra{\rm out}_{\eta \zeta}  \right]^2} 
\left[ 
1 + 2 \Delta (i \mathcal{M}_{\alpha \beta \alpha \beta} + {\rm h.c.}) - \Delta^2 (\mathcal{M}_{\alpha \beta \alpha \beta} - \mathcal{M}_{\alpha \beta \alpha \beta}^{*})^2 \right. \\ & \left. \quad 
+ 2 \frac{E_{p_{b}}}{E_{p_{a}}} \Delta^2 \sum_{\delta} \mathcal{M}_{\alpha \beta \alpha \delta} \mathcal{M}_{\alpha \beta \alpha \delta}^{*}
+ \frac{E_{p_{b}}}{E_{p_{a}}} \Delta^3 \sum_{\gamma, \eta} (i \mathcal{M}_{\alpha \beta \gamma \beta} + {\rm h.c.}) \left( \sum_{\delta} \mathcal{M}_{\alpha \beta \eta \delta'}^{*} \mathcal{M}_{\alpha \beta \gamma \delta'} \right) \right. \\ & \left.  \quad 
+ \frac{E_{p_{b}}}{E_{p_{a}}} \Delta^3 \sum_{\gamma, \eta} (i \mathcal{M}_{\alpha \beta \eta \beta} + {\rm h.c.}) \left(\sum_{\delta} \mathcal{M}_{\alpha \beta \gamma \delta} \mathcal{M}_{\alpha \beta \eta \delta}^{*} \right) + O(\Delta^4)
\right] \,.
\end{aligned}
\end{align}
The factor $\Delta$ can be expressed by $\Delta \equiv T/(sV)$, where $T$ is the total time, $V$ is the volume, and $s$ is the center-of-mass energy squared. Since $\Delta \ll 1$ holds~\cite{Srednicki:2007qs, Kowalska:2024kbs}, we can apply a series expansion in terms of $\Delta$. In our analysis, we consider only the leading contribution.
On the other hand, the denominator in Eq.\,\eqref{eq:trrhoA} can be also obtained by using Eq.\,\eqref{eq:rhoAinte} as 
\begin{align}
\label{eq:fulltrace_app1}
\begin{aligned}
&{\rm tr}_{AB}  \left[ \ket{\rm out}_{\gamma \delta} \bra{\rm out}_{\eta \zeta} \right] \\
&= \sum_{ \gamma, \delta, \eta, \zeta, \gamma', \delta' }\iint d^3 \tilde{p}_{1} d^3 \tilde{p}_{2} \braket{ \mathbf{p}_{1}, \gamma'; \mathbf{p}_{2}, \delta'| {\rm out}}_{\gamma \delta} \braket{ {\rm out} |_{\eta \zeta}^{} \, \mathbf{p}_{1}, \gamma'; \mathbf{p}_{2}, \delta'}  \\
&= 1 + \Delta \left[ i \mathcal{M}_{\alpha \beta \alpha \beta} (p_{a}, p_{b} \to p_{a}, p_{b}) + {\rm h.c.} \right] \\
& \quad + \Delta \sum_{\gamma , \delta} \int d\Pi_{2} \, \mathcal{M}_{\alpha \beta \gamma \delta} (p_{a}, p_{b} \to p_{1}, p_{2}) \mathcal{M}_{\alpha \beta \gamma \delta}^{*} (p_{a}, p_{b} \to p_{1}, p_{2}) \,.
\end{aligned}
\end{align}
Combining Eq.\,\eqref{eq:trrhoA} with Eq.\,\eqref{eq:fulltrace_app1}, we obtain the
\begin{align}
\label{eq:fulltrace}
{\rm tr}_{A}  \left[ (\rho_{\alpha \beta}^{A})^2 \right] \simeq 1 - 2 \Delta \sum_{\gamma , \delta} \int d\Pi_{2} \, \left| \mathcal{M}_{\alpha \beta \gamma \delta} (p_{a}, p_{b} \to p_{1}, p_{2}) \right|^2 + O(\Delta^2)\,.
\end{align}
with the integral over $d{\rm \Pi}_{2}$ is defined as:
\begin{align}
\int d{\rm \Pi}_{2} \equiv \iint d^3 \tilde{p}_{1} d^3 \tilde{p}_{2} (2\pi)^4 \delta^{4}(p_{a} + p_{b} - p_{1} - p_{2}) \,. 
\end{align}

We note that the terms with $O(\mathcal{M})$, which corresponds to forward scattering amplitudes, are canceled due to the second term in Eq.\,\eqref{eq:fulltrace}. 
Finally, the leading contribution in the linear entropy in Eq.\,\eqref{eq:linearEE} is given by 
\begin{align}
\label{eq:EE_AB}
\begin{aligned}
E( \rho_{\alpha \beta}^{A} ) \simeq 2 \Delta \sum_{\gamma, \delta} \int d{\rm \Pi}_{2}
\left| \mathcal{M}_{\alpha \beta \gamma \delta}(p_{a}, p_{b} \to p_{1}, p_{2}) \right|^2 + O(\Delta^2) \,. 
\end{aligned}
\end{align}

Our results are different from those results in Ref.~\cite{Thaler:2024anb}. In  Ref.~\cite{Thaler:2024anb}, they only consider the second term in Eq.\,\eqref{eq:finalstate} due to the use of the post-measurement state. In their context, since $(\rho_{\alpha \beta}^{A})^2 \sim O(\mathcal{M}^4)$, the entanglement entropy is proportional to the fourth power of the scattering amplitudes. In contrast, our formulation results in an entanglement entropy that scales quadratically with the amplitudes, as we include the first term in Eq.\,\eqref{eq:finalstate}. This difference has been analyzed in detail in Ref.~\cite{Kowalska:2024kbs}.

Next we discuss the relationship between entanglement entropy and the cross section. It has been established that the entanglement entropy in scattering processes is related to the corresponding cross section~\cite{Peschanski:2016hgk, Aoude:2024xpx, Low:2024hvn}. The total cross section for the scattering process in Eq.~\eqref{eq:process} is given by: 
\begin{align}
\sigma_{\rm tot} =
\frac{1}{ 2F} \sum_{\gamma, \delta} \int d{\Pi}_{2} 
\left| \mathcal{M}_{\alpha \beta \gamma \delta}(p_{a}, p_{b} \to p_{1}, p_{2}) \right|^2 \,.
\end{align}
with
\begin{align}
F = \sqrt{s^2 - 2 s (m_{\alpha}^2 + m_{\beta}^2) + \left( m_{\alpha}^2 - m_{\beta}^2 \right)^2 } \,. 
\end{align}
Consequently, the entanglement entropy normalized by $\Delta$ is related to the total cross section as:
\begin{align}
\label{eq:CSEE}
\frac{E( \rho_{\alpha \beta}^{A} )}{\Delta}  \simeq 4 F \sigma_{\rm tot} \equiv \overline{E}( \rho_{\alpha \beta}^{A} )\,.
\end{align}
This relationship is consistent with the result in Ref.~\cite{Low:2024hvn}. Eq.\,\eqref{eq:CSEE} suggests that entanglement entropy is not generated if there is no interaction during the scattering process, which is a straightforward interpretation that we confirm numerically in the next section. Furthermore, Eq.~\eqref{eq:CSEE} shows that $\overline{E}( \rho_{\alpha \beta}^{A} )$ approaches a constant value in the high-energy limit, since perturbative behavior requires $\sigma_{\rm tot} \propto 1/s$ when $\sqrt{s} \gg m_{S}, m_{h}$. In addition, the quantity $\overline{E}( \rho_{\alpha \beta}^{A} )$ is dimensionless. Given that $\overline{E}( \rho_{\alpha \beta}^{A} )$ has desirable properties and is independent of $\Delta$, we use $\overline{E}( \rho_{\alpha \beta}^{A} )$ in our numerical calculations instead of $E( \rho_{\alpha \beta}^{A} )$.

We are interested in the change in entanglement entropy induced by the scattering process described in Eq.\,\eqref{eq:process}. A useful measure to quantify this change is the entanglement power, which captures how much entanglement entropy can be generated or reduced by a given operator~\cite{Zanardi:2000zz}. In our calculation, the entanglement power of $\mathcal{S}_f$ is defined as~\cite{Thaler:2024anb}
\begin{align}
\label{eq:ave_Erho}
\mathcal{E}(\mathcal{S}_{f}) 
 &= \frac{1}{N^2} \sum_{\alpha, \, \beta=1}^{N} \overline{E}( \rho_{\alpha \beta}^{A} ) \,,
\end{align}
where the flavor index of the initial state is averaged. 
To gain qualitative insight into the behavior of the entanglement power, we derive an analytical expression for the case $N = 4$: 
\begin{align}
\mathcal{E}(\mathcal{S}_{f}) &\propto \frac{1}{8\pi N^2} \sum_{\alpha, \beta, \gamma, \delta}  \sqrt{1 - \frac{4 m_{S}^2}{s}} |\mathcal{M}_{\alpha \beta \gamma \delta}(s, \theta)|^2 \nonumber \\ 
&\simeq 
\begin{cases}
\frac{2\lambda_{hS}^4 v_{h}^4}{m_{S}^4} \, G \left( \frac{s}{4m_{S}^2}, \frac{m_{h}}{m_{S}} \right) \quad (\lambda_{S} \lesssim \lambda_{hS}) \\
\frac{9 \lambda_{S}^2}{2 \pi} \sqrt{ 1 - \frac{4m_{S}^2}{s}} \quad (\lambda_{S} \gtrsim \lambda_{hS})\\ 
\end{cases}, 
\label{eq:EEanalytic}
\end{align}
where $\theta$ represents the scattering angle, and we evaluate the amplitude at $\theta = \pi/2$ for simplicity. The function $G(x, y)$ is given by:
\begin{align}
G(x,y) = \frac{  8 + 72 x^2 - 12y^2(1 + 3x) + 9 y^4 }{ 16 \pi (y^2 - 4x)^2 (y^2 + 2x -2)^2} \sqrt{ \frac{x-1}{x} } \,. 
\end{align}

\begin{figure}[ht]
\centering
\includegraphics[width=0.5\linewidth]{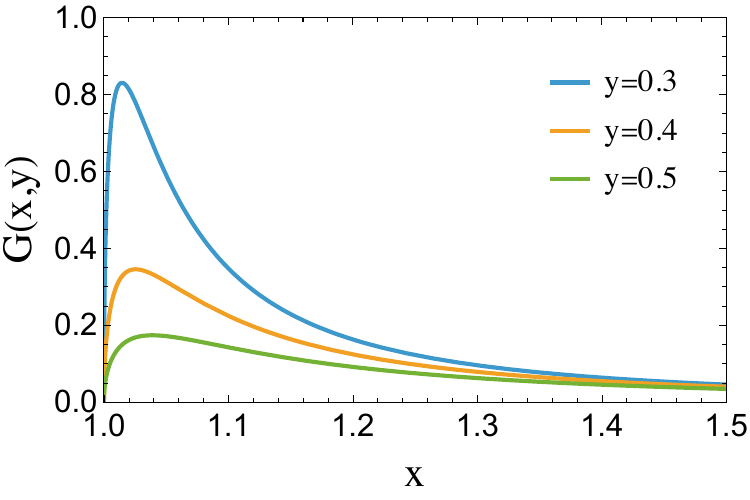}
\caption{ $x$ and $y$ dependence for the function $G(x,y)$. \label{fig:Gxy} }
\end{figure}
The behavior of $G(x,y)$ is illustrated in Fig.\,\ref{fig:Gxy}. When $\lambda_S \lesssim \lambda_{hS}$, Eq.\,\eqref{eq:EEanalytic} shows that entanglement production is enhanced for large $\lambda_{hS}$. In the decoupling limit, where $m_S$ becomes large while $\lambda_{hS}$ is held fixed, the entanglement entropy decreases. This behavior aligns with expectations from the decoupling limit~\cite{Appelquist:1974tg}, and we verify this trend numerically in Section,\ref{sec:numericalresults}.

In contrast, for $\lambda_S \gtrsim \lambda_{hS}$, the entanglement power is dominated by the four-point self-interaction of the $\vec{S}$ field and becomes insensitive to amplitudes involving the $h$ propagator. In the high-energy limit $s \to \infty$, the entanglement power asymptotically approaches $\mathcal{E}(\mathcal{S}_{f})\sim 9 \lambda_{S}^2/(2\pi)$,
as $G(x,y) \to 0$ for $x \to \infty$. This suppression reflects the fact that contributions involving the $h$ propagator vanish as $1/s$ in the high-energy limit.

\section{Numerical results \label{sec:numericalresults}}

\begin{figure}[ht]
\centering
\includegraphics[width=0.48\linewidth]{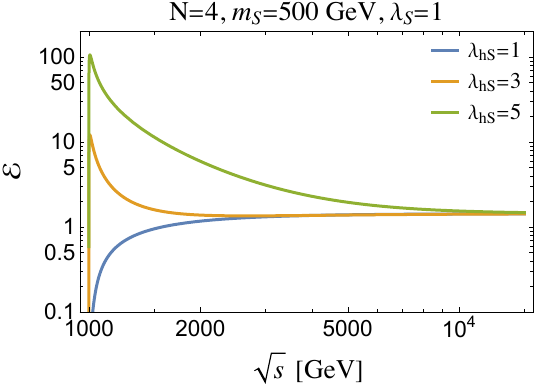}
\includegraphics[width=0.48\linewidth]{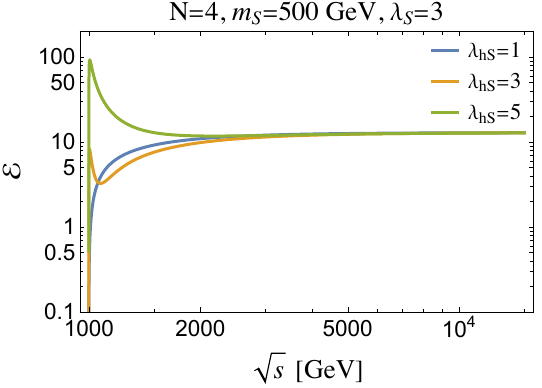}
\caption{
Energy dependence of the entanglement power in the $O(N=4)$ singlet models with $m_{S}=500\,{\rm GeV}$ for fixed $\lambda_{hS}$ and $\lambda_{S}$.
\label{fig:Erho_energy}}
\end{figure}

In this section, we present our numerical results based on the formalism developed in the previous sections. Fig.\,\ref{fig:Erho_energy} shows the energy dependence of the entanglement power for various benchmark point. As $\lambda_{hS}$ increases, the peak value of the entanglement power also grows, indicating that larger Higgs-singlet couplings lead to enhanced entanglement generation. Additionally, at high energies, the entanglement power approaches a constant value specific to each benchmark scenario. This asymptotic behavior is consistent with the analytical results derived in Eq.\,\eqref{eq:EEanalytic} for the large $\sqrt{s}$ limit.

\begin{figure*}[htb]
\centering
\begin{tabular}{cc}
\includegraphics[width=0.45\linewidth]{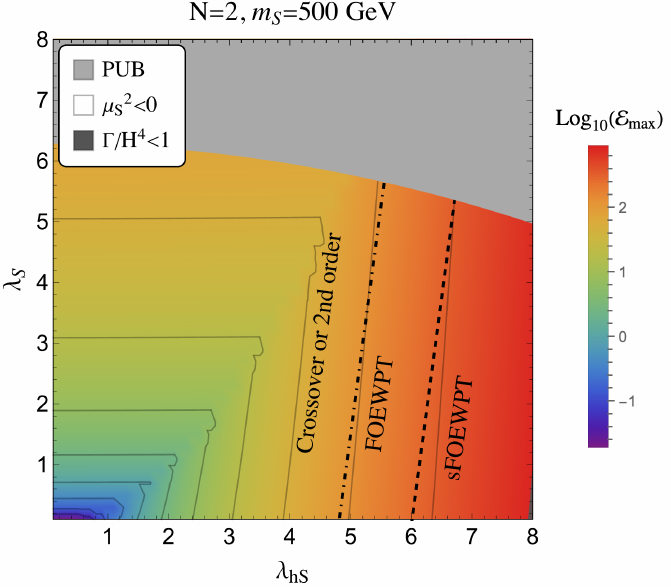}& 
\includegraphics[width=0.45\linewidth]{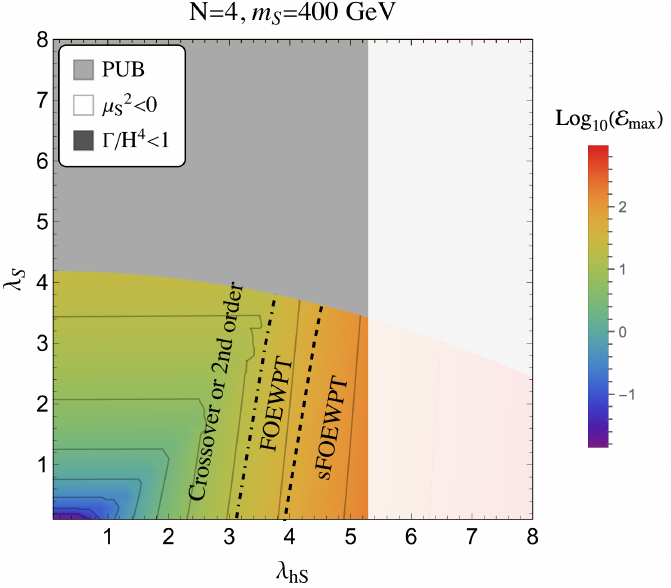}\\ 
\includegraphics[width=0.45\linewidth]{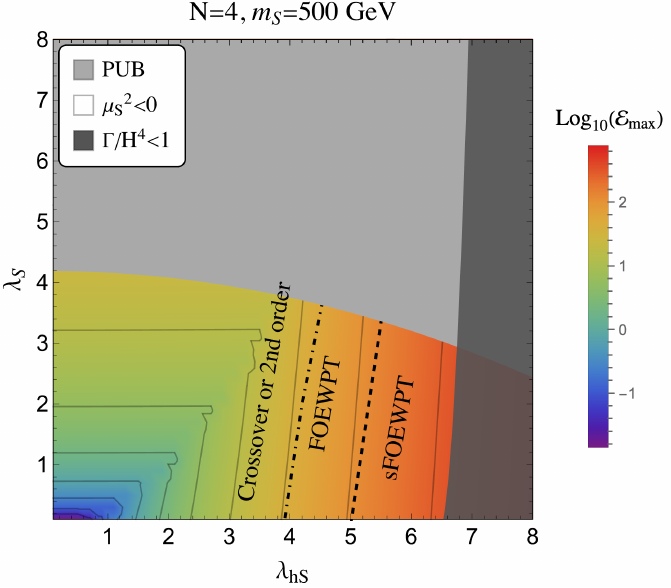}&
\end{tabular}
\caption{
Maximal entanglement entropy in ($\lambda_{hS}, \lambda_{S}$) plane for each benchmark point.
The black dashed and dotted lines indicate the boundary for the sFOEWPT shown in Eq.\,\eqref{eq:vcTc1} and FOEWPT given in Eq.\,\eqref{eq:fopt}, respectively.
The light gray region (PUB) is constrained by the perturbative unitarity bound in Eq.\,\eqref{eq:unitarity}. 
In the white region, the requirement $\mu_{S}^2 >0$ cannot be satisfied. 
The dark gray region is constrained by the requirement that the phase transition should be completed within the age of the universe~\cite{Turner:1992tz}.
}
\label{fig:EEmax}
\end{figure*}

In order to study on the relationship between entanglement power and the sFOEWPT, we define the maximum entanglement power $\mathcal{E}_{\rm max}$ in the region $2m_{S} < \sqrt{s} < 10\,{\rm TeV}$ as
\begin{align}
\mathcal{E}_{\rm max} \equiv \max[\mathcal{E}(\mathcal{S}_{f})] \,.
\end{align} 
Figure~\ref{fig:EEmax} illustrates the dependence of the maximal entanglement power on the parameters $\lambda_{hS}$ and $\lambda_S$ for fixed $m_S$ and $N$. The black dashed and dotted lines correspond to the boundaries for a sFOEWPT and a FOEWPT, as defined in Eqs.\,\eqref{eq:vcTc1} and \eqref{eq:fopt}, respectively. The light gray region labeled ``PUB" is excluded by the perturbative unitarity bound given in Eq.\,\eqref{eq:unitarity}, while the white region is disallowed due to the condition $\mu_S^2 \geq 0$ not being satisfied. The dark gray area (labeled $\Gamma/H^4 < 1$) is excluded because the phase transition would not complete within the age of the universe~\cite{Turner:1992tz}.

As shown in Fig.\,\ref{fig:EEmax}, the maximal entanglement power becomes small in regions where both $\lambda_{hS}$ and $\lambda_S$ are small. This behavior is expected from Eq.\,\eqref{eq:CSEE}, since the relevant cross section is suppressed when the couplings are weak. This result is consistent with previous findings~\cite{Chang:2024wrx, Kowalska:2024kbs}. Conversely, large values of $\lambda_{hS}$ lead to an enhanced entanglement power, as also observed in Fig.\,\ref{fig:Erho_energy}. Notably, these same regions favor the realization of a sFOEWPT. This alignment suggests that the principle of maximizing entanglement entropy production could naturally select regions of parameter space that support a sFOEWPT. Interestingly, such large portal couplings can emerge from UV completions involving strong dynamics~\cite{Kanemura:2012hr,Kanemura:2012uy}. Our results may therefore hint at a deeper connection between entanglement-based criteria and the emergence of new physics at high energy scales.

\begin{figure*}[htb]
\begin{center}
\begin{tabular}{cc}
\includegraphics[width=0.45\linewidth]{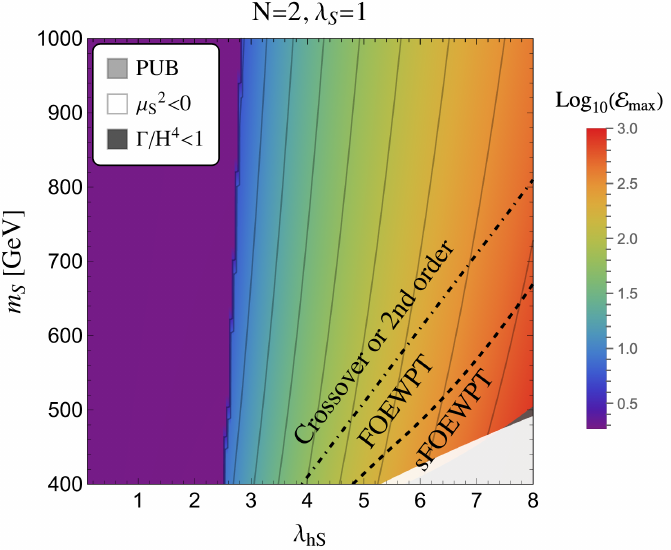} &
\includegraphics[width=0.45\linewidth]{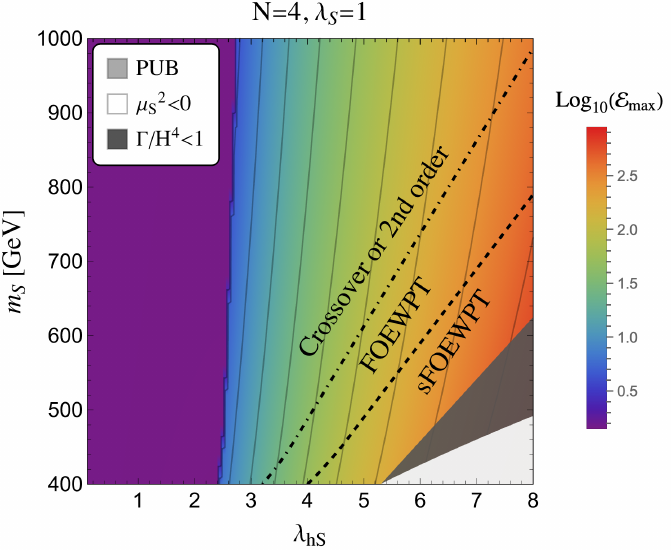} \\
& \includegraphics[width=0.45\linewidth]{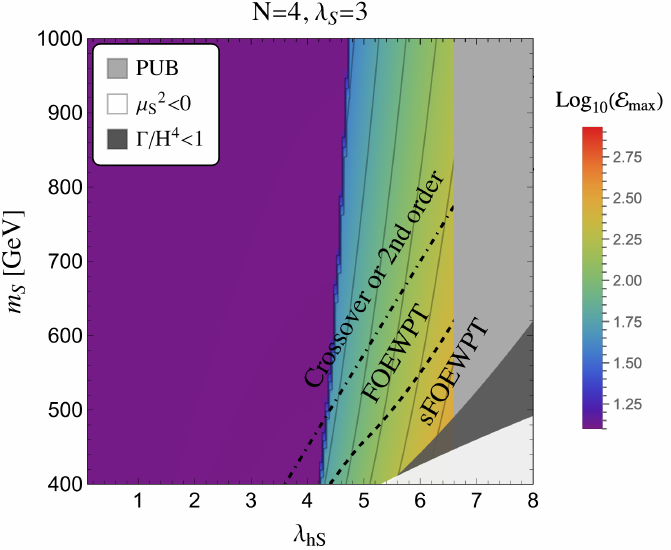} 
\end{tabular}
\caption{
The maximal entanglement power in $(\lambda_{hS}, m_{S})$ plane. 
The gray region is constrained by the perturbative unitarity~\eqref{eq:unitarity}. 
In the lower right region below the black dashed lines, the sFOEWPT can be realized. 
\label{fig:MaxEE_ms_lhs}}
\end{center}
\end{figure*}

Fig.\,\ref{fig:MaxEE_ms_lhs} presents the maximal entanglement power in the $(\lambda_{hS}, m_S)$ parameter plane. We find out that large entanglement power is achieved in the region with small $m_S$ and large $\lambda_{hS}$. This behavior suggests that maximal entanglement production is favored when the singlet field $\vec{S}$ acquires its mass predominantly through the Higgs portal interaction, as described in Eq.\,\eqref{eq:Smass}. In such scenarios, non-decoupling effects become significant, indicating that new physics contributions from light singlet states with strong Higgs interactions play a crucial role in enhancing entanglement.

Moreover, as confirmed in Fig.\,\ref{fig:EEmax}, the sFOEWPT prefers the large entanglement power.
Whereas, in the small $\lambda_{hS}$ region, the parameter dependence disappears. 
The reason is that in such region the maximal entanglement power is always realized in the large $\sqrt{s}$ region as shown in Fig.\,\ref{fig:Erho_energy}. 
Then, the amplitude containing $\lambda_{hS}$ does not contribute to the entanglement power because such amplitude vanishes in the large $\sqrt{s}$ limit due to the inner propagators. 
Therefore, the entanglement power depends only on $\lambda_{s}$ in the large $\sqrt{s}$ limit. 
As a result, the small $\lambda_{h S}$ region in Fig.\,\ref{fig:MaxEE_ms_lhs} does not show the parameter dependence.

\begin{figure*}[ht]
\centering
\includegraphics[width=0.48\linewidth]{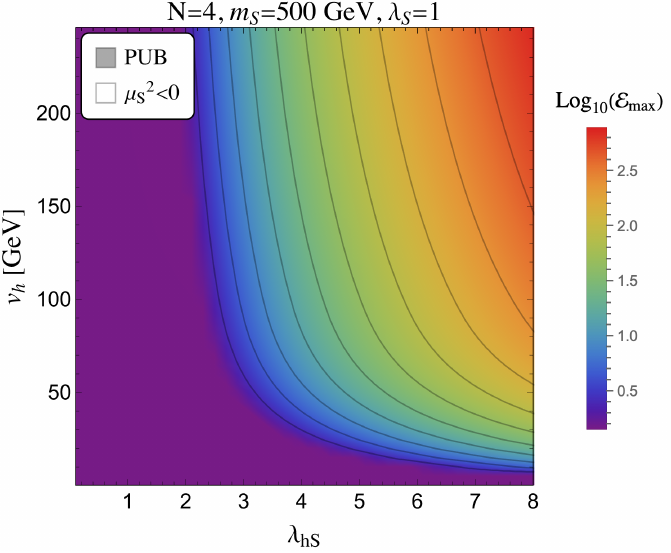}
\includegraphics[width=0.48\linewidth]{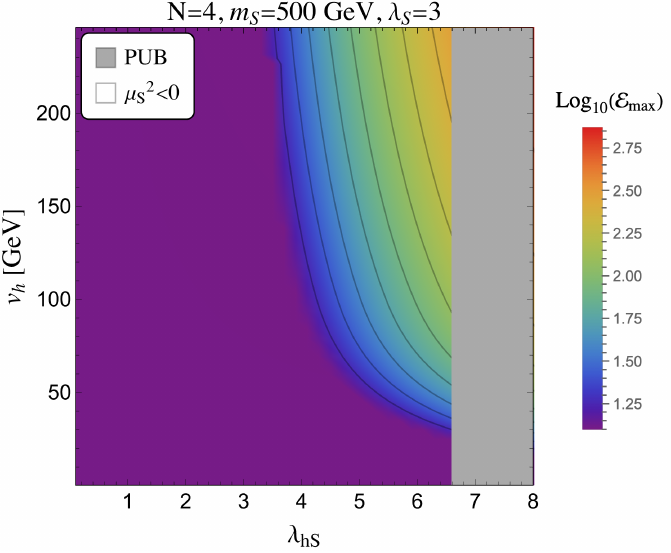}
\caption{
The maximal entanglement power in $(\lambda_{hS}, v_{h})$ plane. 
When the FOEWPT occurs ($\braket{\phi} = 0 \to \braket{\phi} \neq 0 $), the discontinuous production of entanglement entropy occurs. 
The gray region is constrained by the perturbative unitarity bound given in Eq.\,\eqref{eq:unitarity}. 
In the case with $m_{S} =500\,{\rm GeV}$, the condition $\mu_{S}^2 \geq 0$ is always satisfied in the parameter region on which we focus. 
\label{fig:PT_MaxEE_ms_lhs}}
\end{figure*}
Figure~\ref{fig:PT_MaxEE_ms_lhs} shows the maximal entanglement power in the $(\lambda_{hS}, v_h)$ plane. The gray region is excluded by the perturbative unitarity bound. To obtain this result, we vary $v_h$, which effectively mimics the temperature evolution of the Higgs VEV. Therefore, this figure can be interpreted as showing how the entanglement power evolves with temperature. We find that in the large $\lambda_{hS}$ region, the entanglement power increases significantly with $v_h$. As previously shown in Fig.~\ref{fig:EEmax} and Fig.\ref{fig:MaxEE_ms_lhs}, the sFOEWPT typically occurs in such a a region. During a FOEWPT, the Higgs VEV undergoes a sudden jump from $\langle\phi\rangle = 0$ to $\langle\phi\rangle \neq 0$, which leads to a discontinuous change in the maximal entanglement power $\mathcal{E}_{\max}$, which is a clear signature of a FOPT. This behavior aligns well with conventional thermodynamic expectations and suggests that the entanglement power could serve as a useful order parameter for the EWPT.

\begin{figure*}[t]
\centering
\includegraphics[width=0.48\linewidth]{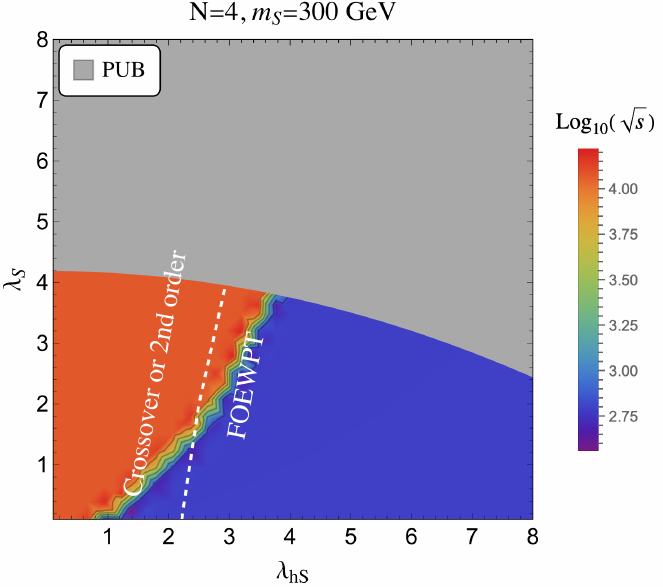}
\includegraphics[width=0.48\linewidth]{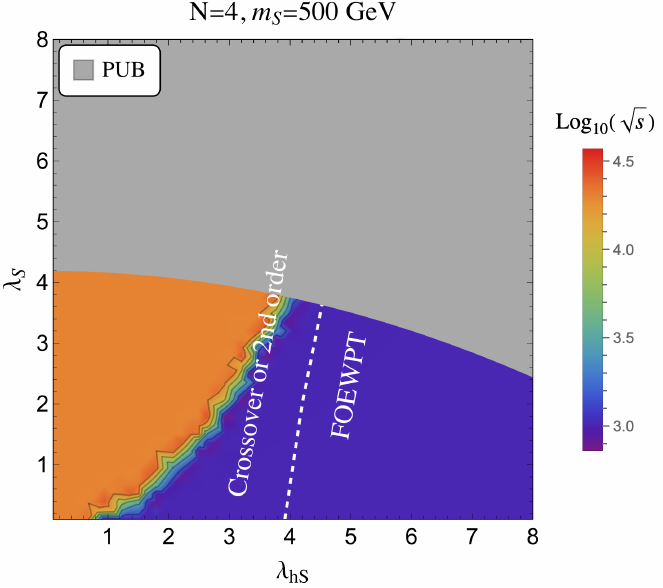}
\caption{
The prediction for the center-of-mass energy $\sqrt{s}\,[\rm GeV]$ at which the maximal entanglement power can be realized.
In the red region, the maximal entanglement power can be realized in the large $\sqrt{s}$ limit. 
In the blue region, we can have a non-trivial maximal entanglement power at a certain small value of $\sqrt{s}$. 
In the right side of the white dashed line, the EWPT can be first order. 
\label{fig:phasediagram}}
\end{figure*}
Figure~\ref{fig:phasediagram} presents the predicted center-of-mass energy $\sqrt{s}$ at which the maximal entanglement power is achieved. In the red region, this maximum occurs at large $\sqrt{s}$, while in the blue region, it is realized at a lower value of $\sqrt{s}$. The white dashed line marks the boundary beyond which the EWPT can be first-order. Notably, in the case with $m_S = 300\,\text{GeV}$, the boundary between the blue and red regions closely aligns with the white dashed line near $\lambda_{hS} \sim 3$. This suggests that, in certain parameter spaces, the energy scale at which entanglement is maximized may provide information about the nature of the EWPT. In such cases, entanglement power could potentially serve as a diagnostic tool for identifying sFOPT. However, as shown in the right panel of Fig.~\ref{fig:phasediagram}, this correlation does not hold universally across all benchmark points. Therefore, while intriguing, this behavior is model-dependent and should not be taken as a general prediction.

\section{Discussions \label{sec:discussion}}

In our study, we focus on maximal entanglement power as a guiding principle in the first place as following reasons.
First, as shown in Fig.\,\ref{fig:Erho_energy}, maximal entanglement power is realized when the coupling $\lambda_{hS}$ becomes large. Importantly, beyond a certain threshold of $\lambda_{hS}$, non-trivial extrema of entanglement appear. This means the condition $d\mathcal{E}/ds = 0$ admits non-trivial solutions at certain small values of $\sqrt{s}$. This indicates that the dynamics of entanglement in our model become qualitatively richer in the strong coupling regime. 

Secondly, this approach finds conceptual support in the context of holographic entanglement entropy. As discussed in the introduction, prior studies have shown that the critical temperature of confinement/deconfinement phase transitions correlates with points where the derivative of the holographic entanglement entropy with respect to the domain scale $\ell$ vanishes~\cite{Klebanov:2007ws, Nishioka:2006gr, Nakagawa:2009jk, Itou:2015cyu}. This parallels our observation that critical features of the theory align with extrema of entanglement. Such a structural similarity suggests that extremal entanglement may serve as a universal signature of phase transitions or critical behavior. Finally, it has been confirmed that the maximal entanglement entropy in helicity states in QED scattering processes predicts the measured weak mixing angle $\sin \theta_{W}$~\cite{Cervera-Lierta:2017tdt}. 
This finding also gives the reason why the requirement of maximal entanglement entropy is plausible. 

In recent, the scattering entanglement entropy in 2HDMs before and after the EW symmetry breaking has been evaluated~\cite{Carena:2025wyh}. 
In Ref.\,\cite{Carena:2025wyh}, it has been shown that the maximal entanglement entropy before the EW symmetry breaking requires the global symmetry $U(2) \times U(2)$.
When the symmetry is gauged to avoid the emergence of Nambu-Goldstone bosons while keeping the maximal entanglement entropy, these gauge sectors should have the $Z_{2}$ symmetry which is related to the existence of dark sector. 
Their results also suggest the importance of maximal entanglement entropy in exploring new physics models.

\section{Conclusions \label{sec:conclusion}}

We have shown that the maximum of the entanglement entropy in the scattering process \eqref{eq:process} favors a large Higgs portal coupling $\lambda_{hS}$.
We have demonstrated that parameter regions with enhanced $\lambda_{hS}$ are capable of realizing the sFOEWPT.
Moreover, we have established that the order of the EWPT may be distinguished by examining the parameter dependence of the entanglement entropy across different setups.
These findings suggest that entanglement entropy may encode meaningful information about the dynamics of electroweak symmetry breaking, offering a novel lens through which to probe fundamental aspects of particle physics.

\section*{Acknowledgment}
J.L. would like to thank Yuichiro Nakai for helpful discussions.
M.T. is grateful to Takumi Kuwahara for fruitful discussions.
The work of J.L., M.T. and J.J.Z. are supported by the National Science Foundation of China under Grant No. 11635001, 11875072, 12235001 and 12475103. 
The work of X.P.W. is supported by the National Science Foundation of China under Grant No. 12375095, and the Fundamental Research Funds for the Central Universities. J.L. and X.P.W. also thank APCTP, Pohang, Korea, for their hospitality during the focus program [APCTP-2025-F01], from which this work greatly benefited. The authors gratefully acknowledge the valuable discussions and insights provided by the members of the China Collaboration of Precision Testing and New Physics.

\bibliographystyle{apsrev4-1}
\bibliography{ref}

\end{document}